\documentclass[twocolumn]{aastex631}

\shorttitle{Reconnection with Advanced Maxwell Solvers}
\shortauthors{Klion et al.}
\graphicspath{{./}{figures/}}

\usepackage{amsmath}
\usepackage{amsbsy}
\usepackage{hyperref}
\DeclareMathOperator{\sech}{sech}
\newcommand{\vect}[1]{\boldsymbol{#1}}
\newcommand{\xc}{x_\mathrm{c}}
\newcommand{\nb}{n_\mathrm{b}}
\newcommand{\nd}{n_\mathrm{d}}
\newcommand{\dilog}{\mathrm{Li}_2}
\newcommand{\thetab}{\theta_\mathrm{b}}

\newcommand{\omegac}{\omega_\mathrm{c}}
\newcommand{\omegacinv}{\omega_\mathrm{c}^{-1}}
\newcommand{\rhoc}{\rho_\mathrm{c}}
\newcommand{\lambdae}{\lambda_\mathrm{e}}
\newcommand{\me}{m_\mathrm{e}}

\begin{document}

\title{Particle-in-Cell Simulations of Relativistic Magnetic Reconnection with Advanced Maxwell Solver Algorithms}

\correspondingauthor{Hannah Klion}
\email{klion@lbl.gov}

\author[0000-0003-2095-4293]{Hannah Klion}
\affiliation{Lawrence Berkeley National Laboratory, 
1 Cyclotron Road, 
Berkeley, CA 94720, USA}

\author[0000-0001-9432-2091]{Revathi Jambunathan}
\affiliation{Lawrence Berkeley National Laboratory, 
1 Cyclotron Road, 
Berkeley, CA 94720, USA}

\author[0000-0003-2406-1273]{Michael E. Rowan}
\affiliation{Advanced Micro Devices, Inc., 
Santa Clara, CA, USA}

\author[0000-0002-9319-4216]{Eloise Yang}
\affiliation{Lawrence Berkeley National Laboratory, 
1 Cyclotron Road, 
Berkeley, CA 94720, USA}

\author[0000-0003-2300-5165]{Donald Willcox}
\affiliation{Lawrence Berkeley National Laboratory, 
1 Cyclotron Road, 
Berkeley, CA 94720, USA}

\author[0000-0002-0040-799X]{Jean-Luc Vay}
\affiliation{Lawrence Berkeley National Laboratory, 
1 Cyclotron Road, 
Berkeley, CA 94720, USA}

\author[0000-0002-3656-9659]{Remi Lehe}
\affiliation{Lawrence Berkeley National Laboratory, 
1 Cyclotron Road, 
Berkeley, CA 94720, USA}

\author[0000-0001-8427-8330]{Andrew Myers}
\affiliation{Lawrence Berkeley National Laboratory, 
1 Cyclotron Road, 
Berkeley, CA 94720, USA}

\author[0000-0003-1943-7141]{Axel Huebl}
\affiliation{Lawrence Berkeley National Laboratory, 
1 Cyclotron Road, 
Berkeley, CA 94720, USA}

\author[0000-0001-8092-1974]{Weiqun Zhang}
\affiliation{Lawrence Berkeley National Laboratory, 
1 Cyclotron Road, 
Berkeley, CA 94720, USA}

\begin{abstract}

Relativistic magnetic reconnection is a non-ideal plasma process that is a source of non-thermal particle acceleration in many high-energy astrophysical systems. Particle-in-cell (PIC) methods are commonly used for simulating reconnection from first principles. While much progress has been made in understanding the physics of reconnection, especially in 2D, the adoption of advanced algorithms and numerical techniques for efficiently modeling such systems has been limited. With the GPU-accelerated PIC code WarpX, we explore the accuracy and potential performance benefits of two advanced Maxwell solver algorithms: a non-standard finite difference scheme (CKC) and an ultrahigh-order pseudo-spectral method (PSATD). We find that for the relativistic reconnection problem, CKC and PSATD qualitatively and quantitatively match the standard Yee-grid finite-difference method. CKC and PSATD both admit a time step that is 40\% longer than Yee, resulting in a ${\sim}40\%$ faster time to solution for CKC, but no performance benefit for PSATD when using a current deposition scheme that satisfies Gauss's law. Relaxing this constraint maintains accuracy and yields a 30\% speedup. Unlike Yee and CKC, PSATD is numerically stable at any time step, allowing for a larger time step than with the finite-difference methods.
We found that increasing the time step 2.4--3 times over the standard Yee step still yields accurate results, but only translates to modest performance improvements over CKC due to the current deposition scheme used with PSATD. 
 Further optimization of this scheme will likely improve the effective performance of PSATD.

\end{abstract}


\section{Introduction} \label{sec:intro}

High-energy radiation is observed from various astrophysical systems such as pulsar wind nebulae and from jets in active galactic nuclei \citep{giannios:10}, X-ray binaries \citep{tetarenko:17}, and gamma-ray bursts \citep{piran:04,kumar:15}.
In particular, pulsars produce high-energy gamma ray flares that evolve too rapidly to be explained by conventional particle acceleration theory \citep{abdo:11,tavani:11}. 
Magnetic reconnection is often invoked to explain the rapid non-thermal particle acceleration and emission in these systems \citep{cerutti:12_particle_accel_crab,nalewajko:15,mckinney:12,petropoulou:19,lyutikov:03,philippov:19}.

During magnetic reconnection, magnetic field energy is converted to particle kinetic energy in the form of both bulk motion and plasma heating. 
For the highly magnetized astrophysical systems, particle acceleration is caused by relativistic magnetic reconnection, where the un-reconnected upstream plasma has a magnetic field energy density many times its enthalpy density. The plasma gains relativistic bulk and thermal velocities, and the particle energy distributions develop non-thermal high-energy power law tails. This population of energized particles are thought to be a source of high-energy emission.

Particle-in-cell (PIC) is a well-established method for studying non-thermal plasma acceleration from first principles \citep{birdsall:91}. Each species in the plasma is modeled with computational particles, which generate currents as they move in the domain. The current is deposited on a spatial grid and is a source term in the Maxwell equations. An electromagnetic field solve step (also referred to as a Maxwell solve) calculates the electric and magnetic fields. The electromagnetic forces are interpolated to the particle positions, and their positions and velocities are advanced accordingly in time. The PIC method therefore fully captures kinetic particle acceleration, as well as the feedback of the accelerated plasma onto the fields.

A number of PIC studies have investigated relativistic reconnection for collisionless electron-positron \citep{zenitani:01,zenitani:07,cerutti:12_particle_accel_crab,sironi:14,nalewajko:15,werner:16} and electron-ion \citep{melzani:14} plasma in two dimensions. Long thin current sheets become unstable to the tearing mode instability, resulting in the formation and mergers of chains of trapped plasma, called plasmoids. Simulations have also indicated that reconnection progresses at a rate of approximately 0.1 in such systems, normalized to the reconnecting magnetic field and the Alfv\'{e}n velocity \citep{guo:15,cassak:17,werner:18}. \citet{cerutti:14_recon_rr} investigated the dispersion relations of the tearing mode (in 2D) and the drift-kink mode that develops in 3D. The fastest-growing tearing mode in simulations has been shown to agree well with analytical expectations \citep{zenitani:07}. Recent work with PIC simulations has focused on the mechanisms underpinning the onset of reconnection  
 and phases of particle energization \citep{guo:19,hakobyan:21,sironi:22}. The particle energy spectra that result from reconnection show hard power laws that extend to high energies \citep{werner:16,werner:18,guo:15,hakobyan:21,petropoulou:19}.
These spectra can then be combined with radiation models to predict observational signatures of reconnection in astrophysical systems \citep{cerutti:13, nalewajko:18}.

While substantial progress has been made in understanding particle acceleration in 2D and, more recently, 3D \citep[e.g.][]{guo:15,zhang:21,schoeffler:23}, algorithmic and computational innovation in PIC simulations of such systems has been limited. Virtually all studies employ a finite-difference time-domain Maxwell solver with a staggered Yee grid \citep{yee:66} (sometimes called FDTD), and only a few have explored the advantages of GPU acceleration for astrophysical PIC simulations \citep{bussmann:13, chien:20, xiong:23}. The Yee approach is second-order in both space and time.
For certain plasma systems, the numerical dispersion inherent to the method can lead to significant errors. With the Yee solver using a time step at the Courant limit, the numerical dispersion error is maximal along the axes and zero along the principal diagonals of the cells.
To obtain solutions with less dispersion, we need alternate solvers, eventually based on higher-order methods. Cole and K\"{a}rkk\"{a}innen 
proposed a non-standard finite-difference approach to mitigate the effects of numerical dispersion along the principal axes when using the time step at the Courant limit \citep{cole:97,cole:02,karkkainen:06}, which \citet{cowan:13} extended to non-cubic cells. This combination is known as the Cole-K\"{a}rkk\"{a}innen-Cowan (CKC) scheme. While numerical dispersion can be suppressed with CKC along the main axes, it remains at other angles.

Higher-order methods, including Fourier-based spectral methods, can be used to reduce dispersion even further. 
Pseudo-Spectral Analytical Time Domain \citep[PSATD,][]{haber:73,vay:13} is one such method, which enables arbitrary-order accuracy that can be set at runtime. Since they are finite-difference schemes, Yee and CKC are only numerically stable when the time step is below a value set by the Courant limit (see Section~\ref{sec:pic_methods}); on the other hand, PSATD is based on analytical integration in Fourier space and has no such constraint. In this paper, we compare the performance and accuracy of the two non-standard approaches, CKC and PSTAD, for relativistic reconnection with the widely-adopted Yee scheme. 
While PSATD does not have a Courant \emph{stability} limit on the time step with regard to the Maxwell solve, a too-large time step may still reduce the accuracy of the simulation (as particles traveling close to the speed of light may travel over a cell size in a single time step). We therefore explore the performance and accuracy of PSATD with time steps above the light travel time across a cell.

\defcitealias{fedeli:22}{Fedeli, Huebl et al. 2022}
The PIC algorithm captures reconnection physics accurately from first principles, but can be computationally expensive, especially when performing high-resolution simulations with higher-order particle shape factors.
Graphics processing units (GPUs) can offer remarkable acceleration over conventional CPU architectures for a number of scientific applications, including PIC \citep{bussmann:13, germaschewski:16,chien:20, vay:20,myers:21}.
We use the GPU-accelerated electromagnetic PIC code, WarpX \citep{vay:20,myers:21}. It has excellent full-machine scaling at leadership-class computing facilities, including Summit and Perlmutter (NVIDIA GPUs) and the world's first reported exascale machine, Frontier (AMD GPUs) \citepalias{fedeli:22}.
The code is built on the AMReX \citep{zhang:19} framework, which supports MPI+$X$ parallelism, where MPI enables inter-rank communications, and $X$ corresponds to an interface such as OpenMP, CUDA, HIP, or SYCL for parallel programming on multi-core CPUs or GPUs.  PIConGPU \citep{huebl:19}, VPIC 2.0 \citep{bird:22}, and the Plasma Simulation code (PSC, \citealt{germaschewski:16}) also employ similar strategies that enable performance portability and allow scaling to multiple GPU nodes. Non-relativistic magnetic reconnection has been used as a comparison case to validate multiple GPU-accelerated PIC codes, including PSC, sputniPIC \citep{chien:20} which can make use of a single node with multiple GPUs, and a CUDA Fortran single-GPU PIC code \citep{xiong:23}.

In this paper, we perform first-of-their-kind 2D GPU simulations of relativistic reconnection with the advanced Maxwell solvers CKC and PSATD. Since these have never before been used for relativistic reconnection, we validate our results by comparing against simulations that use the conventional Yee solver, which has been well studied for 2D systems. In particular, we focus on the evolution of the current sheet structures, the particle-field energy balance, particle energy spectrum, and reconnection rate. We investigate the accuracy-based constraints on PSATD time steps by parametrizing the time step relative to the standard Courant limit of the finite-difference simulations. 
As with the simulations with different solvers, we compare our results with different time steps to the baseline Yee simulations. 
For the same cell size, the Courant limit for CKC admits a longer time step than Yee. PSATD is unconditionally stable with limitations on accuracy that may be imposed by other time-integration algorithms in the PIC simulation. Both CKC and PSATD may allow for faster simulations, so we compare the time to solution for these advanced solvers and assess the performance gains of a large time step with PSATD while holding the cell size constant.

The rest of the paper is organized as follows: In Section~\ref{sec:sim_setup} we describe our simulations, including details about the initial configuration of the current sheets, the perturbation to trigger reconnection, and the algorithmic and numerical parameters. In Section~\ref{sec:solver_results}, we discuss the accuracy and performance results from using different Maxwell solvers. In Section~\ref{sec:cfl_results} we present the results from increasing the time step past the Courant limit with PSATD. We summarize and discuss future directions for our work in Section~\ref{sec:conclusion}.

\begin{figure}
\begin{center}
\includegraphics{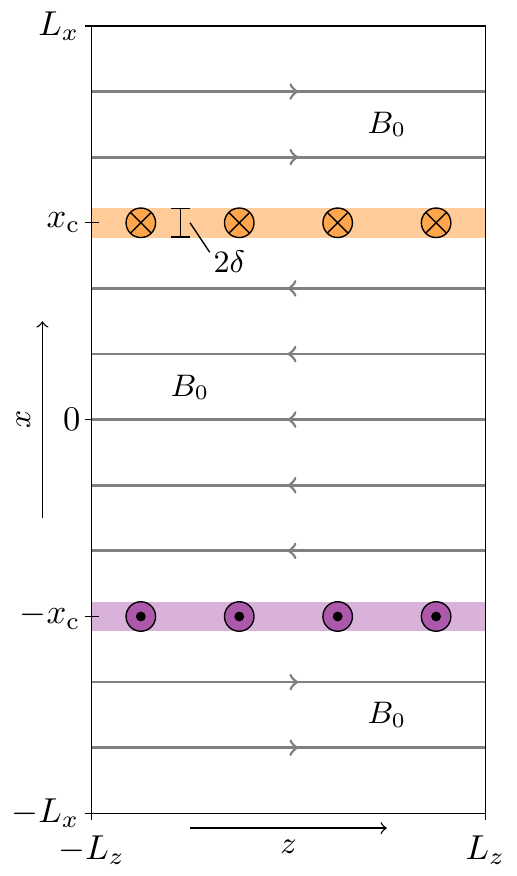}
\end{center}
\caption{\label{fig:cartoon} Diagram of the equilibrium double Harris sheet, before the application of the perturbation. The current sheets are centered at $x =\pm \xc$ and have half-widths of $\delta$ in the x-direction. They extend infinitely in the y-z plane. The current at $x=\xc$ ($x=-\xc$) is into (out of) the plane. The upstream magnetic field has magnitude $B_0$. See equations \ref{eq:nd}, \ref{eq:beta}, and \ref{eq:bz} for expressions for the number densities, particle velocities, and magnetic fields.}
\end{figure}

\section{Simulation Setup} \label{sec:sim_setup}
\subsection{Harris Sheets}
\label{sec:harris}

The simulations are initialized with two pair-plasma Harris current sheets \citep{harris:62} in equilibrium. This section describes the configuration, which is summarized in the diagram in Figure~\ref{fig:cartoon}. Full details of the derivation are in Appendix~\ref{app:magnetic_field}, and summaries of important parameters can be found in Table~\ref{tab:params} (scaled units) and Table~\ref{tab:params_numerical} (SI units).  Input files to replicate the simulations and analysis are available online\footnote{\url{https://doi.org/10.5281/zenodo.7847375}}.

Far from the current sheets, in the upstream, the magnetic field is $\vect{B} = \pm B_0 \vect{\hat{z}}$. Its sign changes at the current sheets. Our principal unit of time will be the inverse upstream electron gyrofrequency, $\omegacinv \equiv \me/(e B_0)$, where $c$ is the speed of light, $e$ is the elementary charge, and $\me$ is the electron mass. Our base unit of length will be $\rhoc = c\omegacinv$, which is the nominal relativistic Larmor radius. 
The current sheets are chosen to have half-width $\delta = 12.15\,\rhoc$, and extend in the y-z plane. They are centered at $x = \pm\xc \equiv \pm L_x / 2$, where $L_x$ is the half-width of the domain in the x-direction, which spans the interval $[-L_x,L_x]$.

We establish spatial distributions of number density $n(x)$ and bulk velocity $\vect{\beta}(x)$:
\begin{align}
    n(x) &= \nb + (\nd - \nb)\left( \sech\frac{x+\xc}{\delta}+\sech\frac{x-\xc}{\delta}\right)\label{eq:nd}\\
    \vect\beta(x) &= \beta(x) \vect{\hat{y}}\\ &= \beta_0 \left( \sech\frac{x+\xc}{\delta}-\sech\frac{x-\xc}{\delta}\right) \vect{\hat{y}} \label{eq:beta}.
\end{align}
Each species has number density $n(x)$. Positrons have velocity $\vect\beta(x)$, and electrons have velocity $-\vect\beta(x)$. The number density in the upstream region is $\nb$, and $\nd$ in the current sheet. The bulk velocity at the center of the current sheet is $\beta_0$, and $\vect{\hat{y}}$ is the unit vector parallel to the y-axis.
From Ampere's Law, $\mathbb{\nabla \times} \vect{B} = \mu_0 \vect J$:
\begin{align}
-\frac{\partial}{\partial x} B_z &= 2 \mu_0 e n(x)\beta(x) c
\label{eq:amplaw}
\end{align}
where $\vect{B}$ is the magnetic field, $\vect{J}$ is the current density, and $\mu_0$ is the vacuum permeability.
The factor of 2 in front of $J_y$ is due to there being two species, positrons and electrons, that contribute to the total current density. Except where otherwise indicated, all quantities are given in the observer frame.
We solve equation \ref{eq:amplaw} for $B_z(x)$ with the boundary conditions
\begin{align}
    B_z(\infty) = -B_z(0) = B_0. \label{eq:bcs}
\end{align}

This gives
\begin{align}
    B_z&(x) = - \frac{ 2 B_0 }{\left(\frac{\pi}{2} + \frac{\nd}{\nb}-1\right)} \times \nonumber\\ 
    &\left( \arctan	\tanh\frac{x+\xc}{2\delta}- \arctan	\tanh\frac{x-\xc}{2\delta} - \frac{\pi}{4}\right.\nonumber\\ 
    & + \left.\frac{1}{2}\left[\frac{\nd}{\nb}-1\right]\left[\tanh\frac{x+\xc}{\delta}- 	\tanh\frac{x-\xc}{\delta} - 1\right]\right), \label{eq:bz}
\end{align}\\
where 
\begin{align}
   B_0 &= 2\mu_0 e \nb \beta_0 c \delta \left(\frac{\pi}{2}+\frac{\nd}{\nb}-1\right).\label{eq:B0}
\end{align}

\begin{deluxetable}{ccc}
\tablecaption{Physical parameters and symbols common to all of our simulations. Quantities marked with $^*$ are freely chosen; others are derived. \label{tab:params}}
\tablehead{Parameter & Symbol & Value}
\startdata
%Background Larmor radius & $\rhoc$ & ---\\
%Background Larmor frequency & $\omegac$ & $c/\rhoc$\\
Background (cold) magnetization$^*$ & $\sigma$ & $30$ \\
Background temperature$^*$ & $\theta_\mathrm{b}$ & $0.15$ \\
Current sheet half-width$^*$ & $\delta$ & $12.15 \,\rhoc$\\
Current sheet skin depth & $\lambdae$ & $2.45\,\rhoc$\\
Current sheet overdensity factor$^*$ & $\nd/\nb$ & $5$\\
%Background number density & $\nb$ & --- \\
Current sheet velocity & $\beta_0$ & $0.22\, c$  \\
Current sheet temperature & $\theta_\mathrm{d}$ & $1.57$ \\
Domain half-width (x)$^*$ & $L_x$ & $2195\,\rhoc$\\
Domain half-width (z)$^*$ & $L_z$ & $1058\,\rhoc$\\
\enddata
\end{deluxetable}

We choose the upstream ``cold'' magnetization $\sigma \equiv B_0^2 / (\mu_0 \nb \me c^2) = 30$. This is somewhat different from the typical relativistic ``hot'' magnetization $\sigma_\mathrm{h} = B_0^2 / (\mu_0 h)$, with $h$ the relativistic enthalpy density. The temperature in the upstream will be very mildly relativistic, with $h \lesssim 1.5 \nb \me c^2$, so the hot magnetization is around 20, and $\sigma \sim \sigma_\mathrm{h}$. Based on either definition of magnetization, reconnection will proceed in the highly relativistic regime.

We choose the current sheet overdensity to be a factor of five, such that $\nd = 5 \nb$. The skin depth in the current sheet is, by definition,
\begin{equation}
    \lambdae \equiv \frac{c}{\omega_\mathrm{p}} = c \sqrt{\frac{\me \epsilon_0}{\nd e^2}},
\end{equation}
where $\omega_\mathrm{p}$ is the plasma frequency in the current sheet. From the quantities fixed thus far, $\lambdae = 2.45\rhoc$. 
Combining equation \ref{eq:B0} with expressions for $\sigma$ and $\lambdae$ gives an expression for the velocity at the center of the current sheet:
\begin{equation}
    \beta_0 = \frac{1}{\left(\frac{\pi}{2}+\frac{\nd}{\nb}-1\right)}\frac{\lambdae\sqrt{\frac{\nd}{\nb}\sigma}}{2\delta} = 0.22\, c. \label{eq:beta0}
\end{equation}

We calculate the temperature profile from pressure balance in the x-direction in the observer's (unprimed) frame:
\begin{equation}
\label{eq:pressure_balance}
    P_\mathrm{gas}(x) + P_\mathrm{mag}(x) = C
\end{equation}
where $C$ is a constant.
The gas pressure $P_\mathrm{gas} = T^{xx}$ where $T^{\mu\nu}$ is the stress-energy tensor. In the fluid (primed) frame, $T'^{xx} = 2 n' \theta m c^2$ where $\theta = k_\mathrm{B}T/(\me c^2)$ is the dimensionless temperature, and $n'$ is the number density in the fluid frame. The fluid bulk velocity is in the y direction, so $T'^{xx} = T^{xx}$. The observer-frame number density is $n = \gamma n'$, where $\gamma = 1/\sqrt{1-\beta^2}$. Substituting,
\begin{align}
    P_\mathrm{gas} = T^{xx}&= 2 n m c^2 \theta \gamma^{-1}\\
    &= 2 n m c^2 \theta \sqrt{1-\beta^2}.
\end{align}

The magnetic pressure,
\begin{equation}
P_\mathrm{mag} = \frac{B^2}{2 \mu_0},
\end{equation}
does not need to be transformed since $\vect{B}$ is already in the observer's frame.

We now have
\begin{equation}
    2 n m c^2 \theta \sqrt{1-\beta^2} + \frac{B^2}{2 \mu_0} = C.
\end{equation}
Far from the current sheets,
\begin{align}
n &= \nb\\
\theta &= \thetab = \sigma / \eta\\
\beta &= 0\\
B &= B_0,
\end{align}
where we have chosen $\eta \equiv 200$, so $\theta_\mathrm{b} = 0.15$.
Solving for $\theta(x)$:
\begin{equation}
\label{eq:theta}
    \theta(x) = \frac{\sigma}{4} \frac{(4+\eta)/\eta- [B_z(x)/B_0]^2}{[n(x)/\nb]\sqrt{1-\beta(x)^2}}.
\end{equation}
The temperature at the center of the current sheet is calculated by evaluating equation~\ref{eq:theta} at $x = \pm \xc$, giving $\theta_\mathrm{d} = 1.57$.

Electron-positron pairs are initialized at the start of the simulation; they are arranged such that they are uniformly spaced, and their momenta are initialized by sampling from a Maxwell-J\"{u}ttner distribution at the local temperature $\theta$ \citep{zenitani:15}.

We note that the setup described here is slightly non-standard. We have used spatially-varying distributions of number density (equation~\ref{eq:nd}), bulk velocity (equation~\ref{eq:beta}), and temperature (equation~\ref{eq:theta}) to represent both the upstream plasma and current sheets. More commonly, the current sheet plasma (with fixed hot temperature $\theta_\mathrm{d}$ and drift velocity $\beta_0$) is overlaid on a domain filled with the upstream plasma (temperature $\theta_\mathrm{b}$, number density $n_\mathrm{b}$). Consequently, at initialization, the plasma at the center of the current sheets will have a two-temperature distribution. The number density of the drifting plasma is the only quantity that varies spatially. Ultimately, the resulting current distributions are very similar, as are the induced magnetic fields, and both configurations are in equilibrium. The primary difference is that the values of $\theta_\mathrm{d}$ and $\beta_0$ used here differ from those of \citet{werner:18}, whose Harris sheets may otherwise appear identical to ours.

\subsection{Perturbation}
\label{sec:perturbation}

 We add a perturbation to the equilibrium configuration in order to control the location and number of x-points. We model our perturbation after \citet{werner:18}, who add a one percent sinusoidal perturbation to the vector potential, which decreases the magnetic pressure at the z-axis just outside of each of the current sheets: 
\begin{align}
    A_y = \int &B_z(x) \,\mathrm{d}x \,\times \nonumber\\
    &\left[1- 0.01 \cos^{51}\left(\frac{\pi}{L_z}z\right) \cos^2\left(\frac{\pi(x-x_{c})}{L_x}\right)\right].\label{eq:Ay}
\end{align}
We set the constant of integration equal to zero. 
In most studies that use this perturbation, the cosine in z is not raised to any power. We find that instead raising it to the 51st power narrows the region being perturbed and thus reduces the number of initial X-points in our configuration. This helps to ensure a more consistent comparison between the simulations. However, after the initial phases of reconnection, the form of the perturbation does not affect the results. The exact choice of power is arbitrary, though must be odd in order to maintain the sign of the cosine term.

We initialize the fields in the simulation with $\vect{B} = \nabla \times (A_y \vect{\hat{y}})$. The simulations will therefore all start with $\nabla \cdot \vect{B} = 0$, which will be conserved by all numerical methods used here.

\subsection{Particle-in-Cell Simulations}
\label{sec:pic_methods}

Simulations of relativistic electron-positron pair plasma reconnection were performed using the electromagnetic PIC code WarpX \citep{vay:20,myers:21}.
The two-dimensional domain is discretized with a uniform grid with a cell size of $\Delta x = \Delta z = \lambdae/4$. Both the skin depth in the current sheet and upstream Larmor radius $\rhoc = 1.63\Delta x$ will be resolved. Near the end of the simulation, the expectation is that energized particles will have Larmor radii of $\rho_\mathrm{c,f} = \sigma\rhoc$. Since the grid resolves the smaller initial Larmor scale, it will continue to resolve it throughout the simulation as it grows.

The simulations were initialized with 64 particles per species per cell, evenly spaced in both directions. At this particle density, the overall results are converged, and statistical noise is  not significant. The grid extends to $[-L_x,L_x] \times [-L_z,L_z]$, with $L_x = 2195\,\rhoc = 73.2\,\rho_\mathrm{c,f}$ and $L_z = 1058\,\rhoc = 35.3\,\rho_\mathrm{c,f}$. The simulations have a resolution of $(7168 \times 3456)\,\mathrm{cells}$, and a total of 3.1~billion particles. Periodic boundary conditions are applied on all domain edges. The full domain is ${\sim}70\,\rho_\mathrm{c,f}$ in the smaller dimension, situating our simulations in the ``large-domain'' regime where particle acceleration is expected to be primarily limited by the plasma properties rather than the box size \citep{werner:16}.

The choice of electromagnetic field solver sets the time step. In this paper, three choices are explored: the standard finite-difference time-domain method on a Yee grid (here called ``Yee'') \citep{yee:66}, a non-standard finite-difference Cole-K\"{a}rkk\"{a}inen solver with Cowan coefficients (CKC) \citep{cole:97,cole:02,karkkainen:06,cowan:13}, and a pseudo-spectral analytical time domain (PSATD) method \citep{haber:73,vay:13} with a 16th-order stencil \citep{vincenti:16}. On our uniform 2D grid with square cells, Yee has a maximum stable time step $\Delta t_\mathrm{C}$ set by the Courant-Friedrichs-Lewy (CFL) condition: $\Delta t_\mathrm{C} = \Delta x / (c\sqrt{2})$. CKC admits a larger time step $\Delta t_\mathrm{C} = \Delta x /c$. Unlike the finite-difference schemes, PSATD does not have a theoretical limit on time step since it is unconditionally stable. In that case, by default, WarpX sets the time step to be the smallest cell light-crossing time $\Delta t_\mathrm{C} = \Delta x / c$ as a Courant-like time step. For the remainder of the paper, we define the ``CFL factor'', simply denoted ``CFL'', as a multiplying factor to the CFL time step limit $\Delta t_\mathrm{C}$ of a given Maxwell solver, such that the time step of a simulation is given by $\Delta t = \mbox{CFL}\times \Delta t_\mathrm{C}$.

Two sets of simulations were performed. The first set compares the three electromagnetic field solvers with a time step $\Delta t = 0.95 \Delta t_\mathrm{C}$ (CFL factor of 0.95). The second studies the effect of time steps that exceed $\Delta t_\mathrm{C}$ with PSATD. While there is no formal stability limit on the time step, the accuracy of our simulations can be expected to deteriorate for CFL factors larger than some value to be determined. Using a systematic study of the evolution of the simulations for a range of CFL factors, we will determine if and in what cases such a practical time step ceiling exists. 

By default, each Maxwell solver is paired with a current deposition algorithm that guarantees charge conservation. The Esirkepov deposition scheme \citep{esirkepov:01} is charge-conserving when used with either Yee or CKC \citep{vay:11}. However, Gauss's law is not satisfied when Esirkepov deposition is combined with high-order PSATD. \citet{vay:13} developed a deposition scheme (hereafter referred to as ``Vay'') that preserves $\vect{\nabla} \cdot \vect{E} = \rho / \epsilon_0$ when used with PSATD. Here, $\rho$ is the charge density, so we also refer to algorithmic combinations that satisfy this equation as ``charge-conserving''. We exclusively use Esirkepov with Yee and CKC. When not otherwise specified, PSATD is coupled with the Vay deposition. These three combinations form our main comparison. When measuring the time to solution, we find it instructive to decouple the performance differences from the Maxwell solvers and the current deposition schemes. To do so, we compare the main three simulations against a PSATD simulation with Esirkepov deposition. It happens that for this problem, PSATD+Esirkepov produces a correct result despite violating Gauss's law. Both deposition schemes use cubic splines for the particles, and once on the grid, currents are smoothed with a single-pass bilinear filter.

In all cases, the field gather operation also uses cubic splines. We use a relativistic second-order Boris push to advance the particle positions \citep{boris:70}. 

\begin{figure*}
\begin{center}
\includegraphics{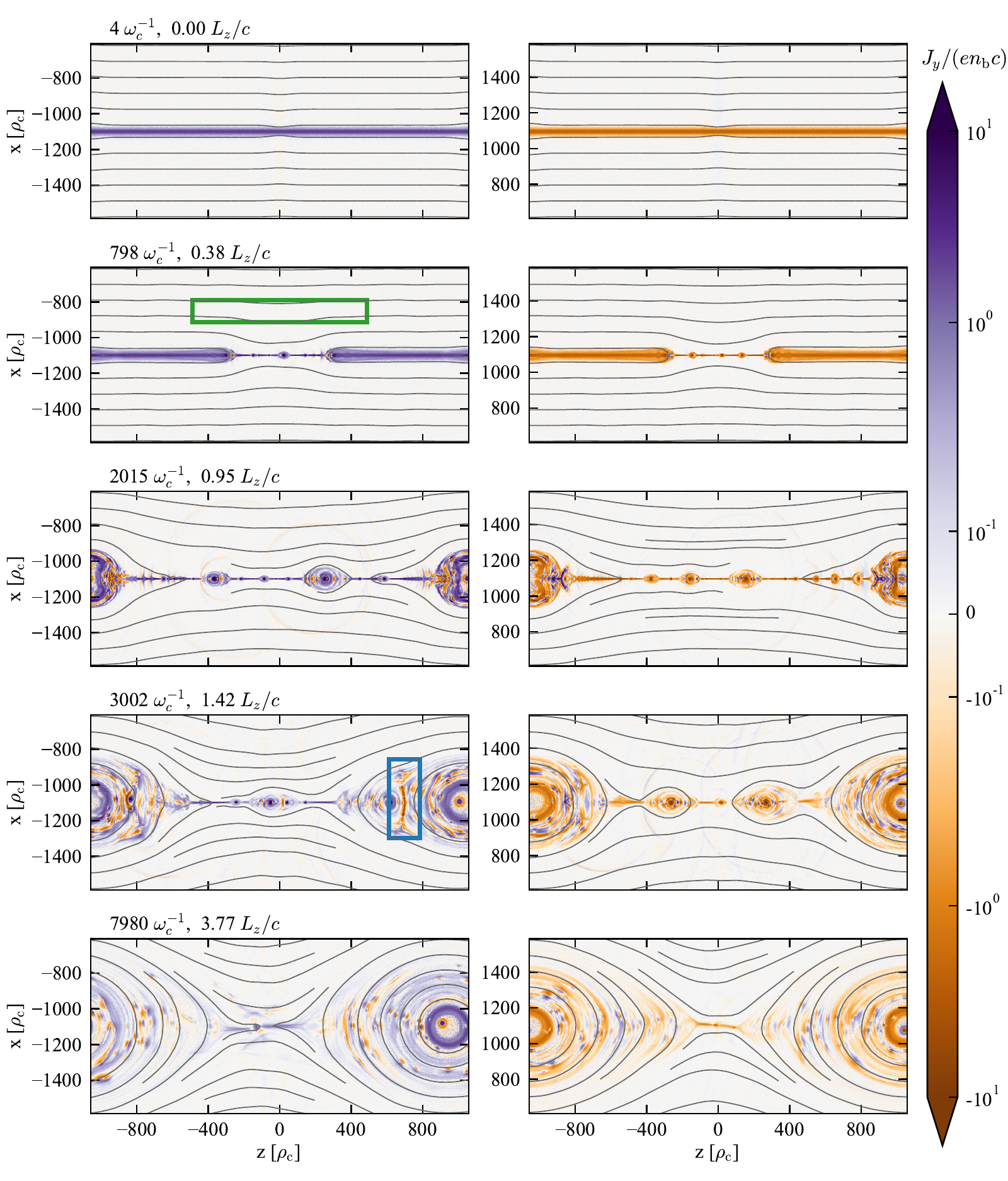}
\caption{\label{fig:Jy_Bz_time} Time evolution of top (left column) and bottom (right column) current and in-plane magnetic field (black lines) in the Yee simulation. The small perturbation to the magnetic fields at $z = 0$ at the initial current sheets leads the current sheet to collapse at that point, thinning it out and causing reconnection to start. The current sheet fragments into plasmoids, which move away from the center and merge with one another, causing secondary reconnection (e.g.\@ region marked with blue box). At the end, there is a single large plasmoid, and reconnection ends. The green box shows the region used to calculate average inflow velocity in Section~\ref{sec:rate}.}
\end{center}
\end{figure*}

\begin{figure}
\begin{center}
\includegraphics{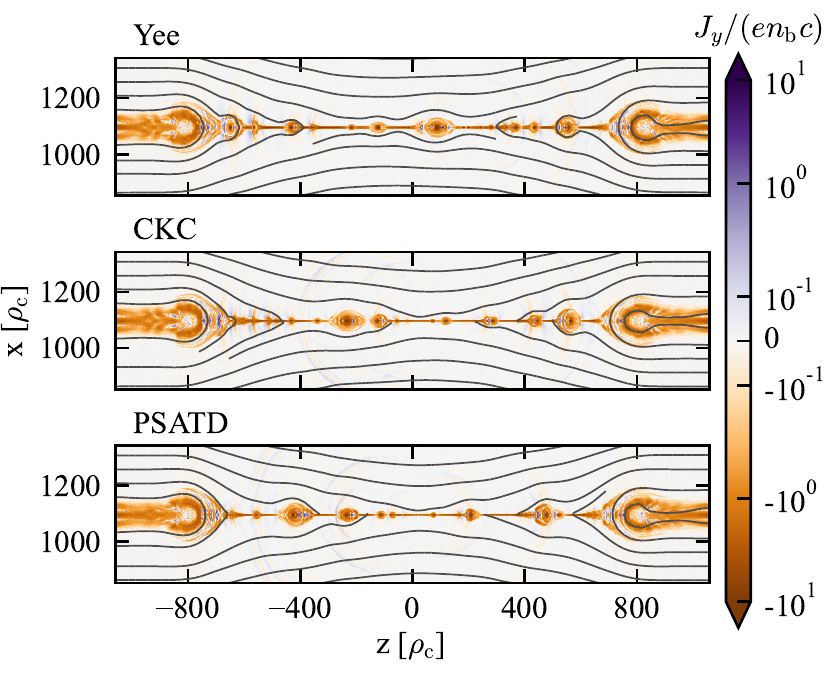}
\caption{\label{fig:Jy_Bz_1470} Comparison between current sheet and magnetic field structure in simulations using different Maxwell solvers. All snapshots are shown at $t=1470\,\omegac^{-1}$. At this phase, all simulations show that the current sheet has fragmented into several small plasmoids, and a single large plasmoid is forming around the edge of the domain. There are small differences between the current and magnetic field structures that are due to the nonlinearity of reconnection. The energy conversion and particle acceleration are similar to one another (see Figures \ref{fig:energy_total_solver} and \ref{fig:power_spectra_solver}).}
\end{center}
\end{figure}

\section{Results: Maxwell Solvers}
\label{sec:solver_results}

\subsection{Energy Conversion and Particle Acceleration}
\label{sec:energy_conversion}

The PIC simulations of reconnecting Harris sheets (described in Section~\ref{sec:sim_setup}) were conducted until reconnection has completed and magnetic energy is no longer being converted to particle kinetic energy, at around $t = 8000$--$9000\,\omegac^{-1} \approx 4\times(2 L_z/c)$. The qualitative current evolution of the Yee simulation is captured in Figure~\ref{fig:Jy_Bz_time}. The other simulations evolve similarly. Shortly after the start of the simulation, both the top and bottom current sheets (left and right columns in Figure~\ref{fig:Jy_Bz_time}) collapse due to the perturbation (equation \ref{eq:Ay}) that lowers the magnetic pressure  just above and below the current sheet (second row in Figure~\ref{fig:Jy_Bz_time}). Quasi-circular structures of trapped plasma and current form, called plasmoids. They inherit their average current from the current sheet where they form. Several magnetic X-points form between the plasmoids in the current sheets. The plasmoids move outward along the current sheet, and occasionally merge, as highlighted in blue box in the fourth row of Figure~\ref{fig:Jy_Bz_time}. This merger creates a new current sheet anti-parallel to the bulk current in the plasmoids, and that extends perpendicularly from the original. This is the site of `secondary reconnection'. At $t \approx 8000\,\omegac^{-1}$, both primary and secondary reconnection have ended, leaving a single plasmoid at $z \approx \pm L_z$ (bottom row in Figure~\ref{fig:Jy_Bz_time}). 

In all three of the Maxwell solvers studied, magnetic reconnection proceeds at approximately the same rate and with the same structures. Figure~\ref{fig:Jy_Bz_1470} compares the plasma structures around the upper current sheet at $t = 1470\,\omegacinv$ for simulations that use the Yee, CKC, and PSATD solvers. Several small plasmoids have formed in each current sheet, with a single larger one forming at the edge of the domain. The exact positions and sizes of the smaller plasmoids differ between the solvers, but the current sheet fragments in a similar way in all cases.

\begin{figure}
\includegraphics{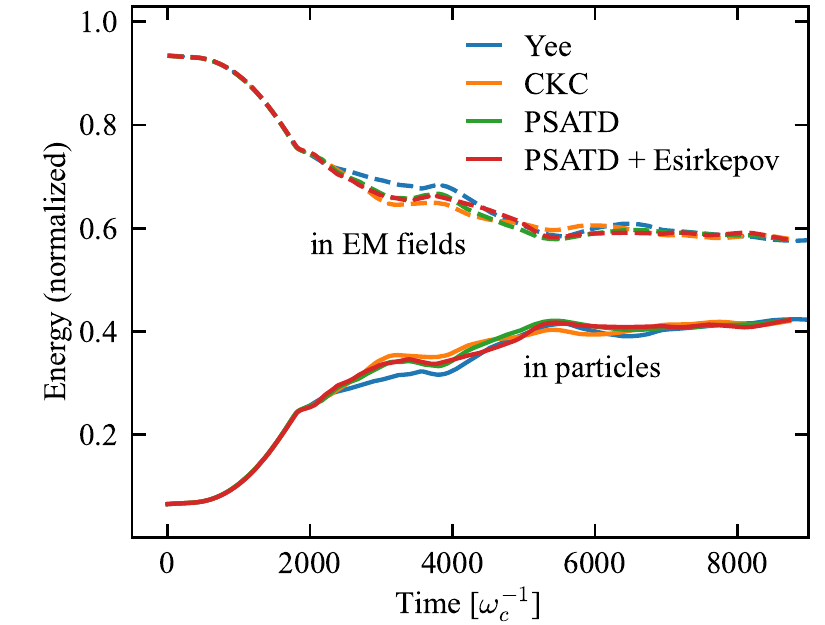}
\caption{\label{fig:energy_total_solver} Evolution of energy balance between electromagnetic fields (dashed lines) and particles (solid lines) during reconnection simulations. The y-axis is normalized to the total energy at the start of the simulation. Results from different Maxwell solvers are shown in different colors (Yee: blue, CKC: orange, PSATD+Vay current deposition: green, PSATD+Esirkepov current deposition: red). The initial ${\sim}1500\,\omegac^{-1}$ of evolution is nearly identical between all solvers, and after that all simulations show similar evolution, ending when about 40\% of the field energy has been converted to particle kinetic energy.}
\end{figure}

Energy conversion and particle acceleration also proceed similarly with all three solvers. When reconnection saturates at around $t=6000\,\omegac^{-1}$, about 40\% of the energy in electromagnetic fields has been converted to particle kinetic energy (Figure~\ref{fig:energy_total_solver}). This includes energy associated with both thermal and bulk motion. Energy conversion proceeds nearly identically for the first $1800\,\omegac^{-1}$, after which there are slight differences between the numerical methods. This initial interval of identical evolution appears to be one of linear growth of the tearing-mode instability. The amplitudes of the fastest-growing spatial Fourier modes are approximately exponential in this interval. At around $t=1800\,\omegacinv$, the exponential growth stops, suggesting the beginning of a non-linear phase of evolution. This non-linearity coincides with the appearance of small but noticeable differences between the energy conversion when using the different solvers, suggesting that this divergence is a consequence of non-linear evolution. PSATD+Esirkepov shows a similar result.

All simulations conserve energy to within one part in 2000, which is well within an acceptable level of energy non-conservation. The CKC simulation loses $4\times10^{-4}$ of the initial total energy, slightly more than the $3.5 \times 10^{-4}$ and $3\times 10^{-4}$ lost by the Yee and PSATD simulations, respectively.

\begin{figure}
\includegraphics{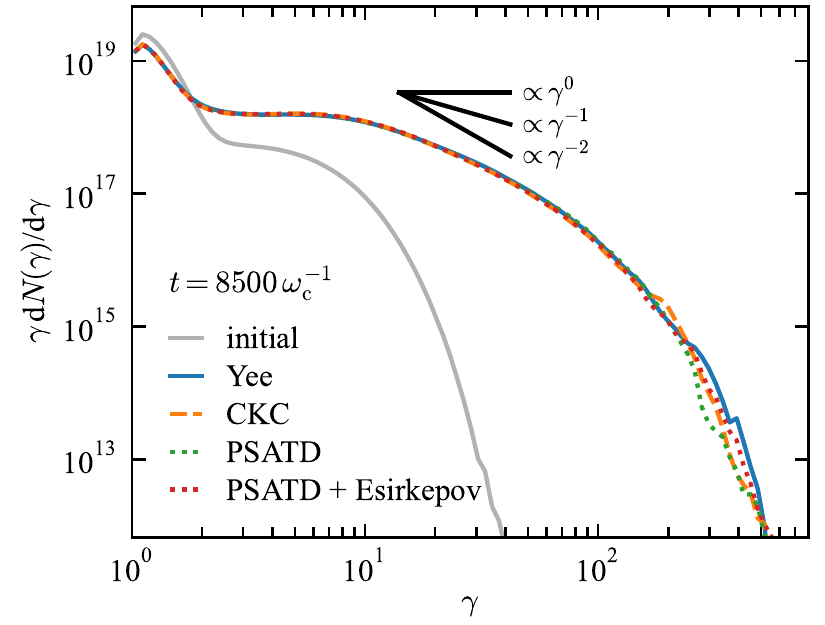}
\caption{\label{fig:power_spectra_solver} Distribution of particle Lorentz factor $\gamma$ weighted by particle energy at the start of the simulation (gray line) and once reconnection has ended ($t = 8500\,\omegac^{-1}$). We compare final distributions for the different Maxwell solvers: Yee (solid blue), CKC (dashed orange), PSATD with Vay current deposition (dotted green), and PSATD with Esirkepov current deposition (dotted red). All three of the Maxwell solvers show very similar particle acceleration, as does the PSATD+Esirkepov combination. The majority of the new particle kinetic energy has gone into particles with $\gamma \sim \sigma$. The spectral slope varies between $0$ and $-2$ (overplotted in black), which roughly matches the ranges found in prior work  \citep{guo:15,werner:16,werner:18}.}
\end{figure}

The three methods and PSATD+Esirkepov also produce quantitatively similar particle acceleration. The highest particle $\gamma$ at the start of the simulation is around $30$, while by the end of reconnection at $t = 8500\,\omegac^{-1}$, the fastest particles have $\gamma = 500$, over ten times higher. The majority of the energy in the domain is in particles with $\gamma \lesssim \sigma$. Prior work has found that $\mathrm{d}N/\mathrm{d}\mathrm{\gamma} \propto \gamma^{-\alpha}$ with $\alpha \approx 1$--$3$ \citep{guo:15,werner:16,werner:18}. This corresponds to power laws with indices between $0$ and $-2$ for the Lorentz factor distribution. Our energy distributions roughly follow this slope (Figure~\ref{fig:power_spectra_solver}).

\begin{figure}
\includegraphics{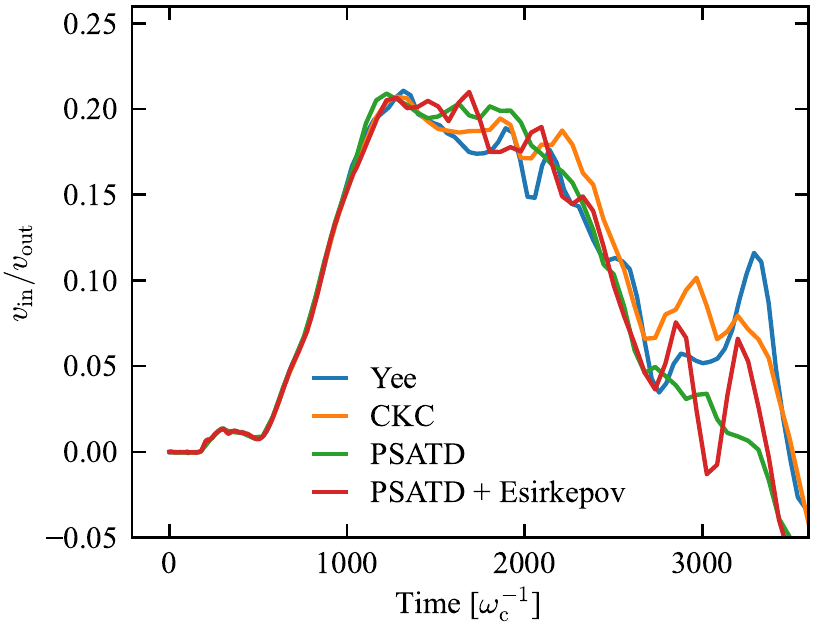}
\caption{\label{fig:recon_rate_solver} Estimated dimensionless reconnection rate $v_\mathrm{in}/v_\mathrm{out}$ for simulations with Yee (blue), CKC (orange), PSATD with Vay current deposition (green), and PSATD+Esirkepov (red). The rate evolves similarly in all cases, peaking at $0.2$ and remaining between $0.15$ and $0.2$ from $t=1000\,\omegac^{-1}$ to $2000\,\omegac^{-1}$. This matches the expected ``universal'' reconnection rate of $0.1$ \citep{comisso:16,liu:17}.}
\end{figure}

\subsection{Reconnection rate}
\label{sec:rate}

The dimensionless reconnection rate
\begin{equation}
    \beta = -\frac{1}{v_\mathrm{A} B_0 L_x} \frac{\mathrm{d}\Phi}{\mathrm{d}t}
\end{equation}
parametrizes much of the linear theory of reconnection \citep[e.g.][]{werner:21}. The Alfv\'{e}n velocity is $v_\mathrm{A}$, and $\Phi$ is the unreconnected flux. Directly measuring the amount of unreconnected flux is difficult, so we instead use the approximation $\beta \approx v_\mathrm{in}/v_\mathrm{out}$, where $v_\mathrm{in}$ is the inflow velocity into the reconnection layer, and $v_\mathrm{out}$ is the terminal exhaust velocity downstream \citep{cassak:17,liu:22}. We calculate $v_\mathrm{in}$ by averaging $|v_x|$ within a region of size $x_R \times z_R = 245\,\rhoc \times 980\,\rhoc = 0.11\,L_x \times 0.92\,L_z$ centered on $(x,z) = (-\xc + x_R, 0)$. This region is marked by the green rectangle in Figure~\ref{fig:Jy_Bz_time}. The measured inflow velocity is relatively insensitive to the choice of $x_R$ and $z_R$. The measurement is also symmetric across the current sheet, i.e.\ moving the box to $x = -\xc-x_R$ does not change the measurement. The measurement is also similar on the other current sheet. For the sake of simplicity, we therefore only show reconnection rates on the +x side of the lower current sheet. The $v_\mathrm{in}$ average only includes cells in the upstream, that is, where more than 70\% of the plasma originated on the same side of the current sheet. Because reconnection mixes plasma across the current sheet, this excludes plasmoids and other reconnection exhaust. The value of this threshold does not strongly affect our results.

We measure the outflow velocity, $v_\mathrm{out}$, by taking the median of the 10 highest cell-averaged z-velocities within $\delta = 12\,\rhoc$ of the center of the current sheet. By $t=1000\,\omegacinv$, this approaches the expected value of $v_\mathrm{A} = c \sigma/\sqrt{\sigma^2+1} \approx c$.

This estimate of the reconnection rate for Yee, PSATD, and CKC simulations is shown in Figure~\ref{fig:recon_rate_solver}. The ratio $v_\mathrm{in}/v_\mathrm{out}$ grows nearly identically in all simulations from 0 to 0.2 at a time of $1200\,\omegacinv$. For the following $1200\,\omegac^{-1}$, the estimate of the rate holds relatively constant between 0.15 and 0.2 before dropping at around $t=2500\,\omegacinv$. This coincides with $2 L_z/c$, the light-crossing time across the current sheet. After this point, the steady-state assumption under which $\beta \approx v_\mathrm{in}/v_\mathrm{out}$ breaks down as the reconnection fronts interfere with one another due to the periodic boundary condition.

The measurement of a rate between 0.15 and 0.2 matches the expectation from prior work. Typically, $\beta$ for non-relativistic reconnection is observed to be around 0.1, with higher rates for relativistic reconnection \citep{blackman:94,guo:15,comisso:16,liu:17}.

\subsection{Performance}
\label{sec:solver_performance}

\begin{deluxetable}{cccc}
\tablecaption{Performance comparison between solvers. The walltime per time step is similar for Yee, CKC, and PSATD+Esirkepov, but 50\% more expensive for PSATD(+Vay). Since PSATD and CKC allow longer time steps, CKC reaches solution 40\% faster, and PSATD(+Vay) is slightly slower than Yee. PSATD+Esirkepov performs comparably to CKC, reflecting that walltime per step is largely governed by the deposition scheme. These numbers are from simulations run to a final time of $t=1470\,\omegac^{-1}$ with dynamic load balancing and no I/O on 21 OLCF Summit nodes (126 GPUs).}
\label{tab:performance}
\tablehead{Algorithmic & Time step & Walltime & Walltime to \\ Options
 & $[\omegacinv]$ & per step [s] & $1470\,\omegac^{-1}$ [s] }
\startdata
Yee & 0.411 & 0.077 & 274.6\\
CKC & 0.581 & 0.077 & 193.5\\
PSATD (+ Vay) & 0.581 & 0.115 & 290.0\\
PSATD+Esirkepov & 0.581 &  0.083 & 209.9\\
\enddata
\end{deluxetable}

Current sheet structures (Figure~\ref{fig:Jy_Bz_1470}), energy conversion (Figure~\ref{fig:energy_total_solver}), particle acceleration (Figure~\ref{fig:power_spectra_solver}), and reconnection rate (Figure~\ref{fig:recon_rate_solver}) are all similar between the Yee, PSATD, and CKC Maxwell solvers. This suggests that reconnection proceeds similarly in all of our simulations, demonstrating that the results produced by the PSATD and CKC solvers for the reconnection problem are comparable to those from the more conventional Yee solver. The PSATD+Esirkepov simulations also produce comparable results, despite the violation of Gauss's law. Adding this simulation to our performance comparison allows us to decouple the performance effects of the Maxwell solvers and current deposition schemes.

In these reconnection simulations, the majority of the runtime (${\sim}60\%$) is spent in the current deposition routine, the performance of which is unaffected by the Maxwell solver. Consequently, the walltime per time step is approximately the same between the Yee, CKC, and PSATD+Esirkepov runs. CKC and PSATD permit a time step that is 40\% longer than in Yee, reducing the time to solution for CKC and PSATD+Esirkepov by 40\% and 30\%, respectively, over the baseline Yee (see Table~\ref{tab:performance}). PSATD+Esirkepov's steps take 10\% longer than Yee(+Esirkepov) or CKC(+Esirkepov), indicating that PSATD field-solve itself has only a small impact on the computational performance. However, PSATD(+Vay) has a 50\% longer time per step due to differences in the current deposition kernels. Thus, in spite of the longer timestep permitted by the PSATD field solver, there is only a slight net increase in the time to solution. The implementation of Esirkepov deposition in WarpX is highly optimized, so it is possible that similar optimization in Vay deposition could make PSATD+Vay a more advantageous combination. This is an area for future work. 

\section{Results: Large Time Steps with PSATD}
\label{sec:cfl_results}

A particular advantage of the PSATD method over either CKC or Yee is that it is not subject to a Courant stability criterion. Consequently, we are able to further increase the time step in the PSATD simulations by raising the CFL factor above 1. If other algorithmic choices are kept the same, then the time per time step is unlikely to change, reducing the time to solution. In this section, we study a sequence of simulations that are identical except for their CFL factors, and therefore time steps. The CFL factors studied range from a baseline of 0.95 to 2.2. We refer to simulations in this sequence as CX.X where `X.X' is the CFL factor. C0.95 is the simulation labeled as `PSATD' in Section~\ref{sec:solver_results}. As discussed in Section~\ref{sec:sim_setup}, we exclusively use the Vay current deposition scheme for the simulations in this sequence. 

\begin{figure}
\includegraphics{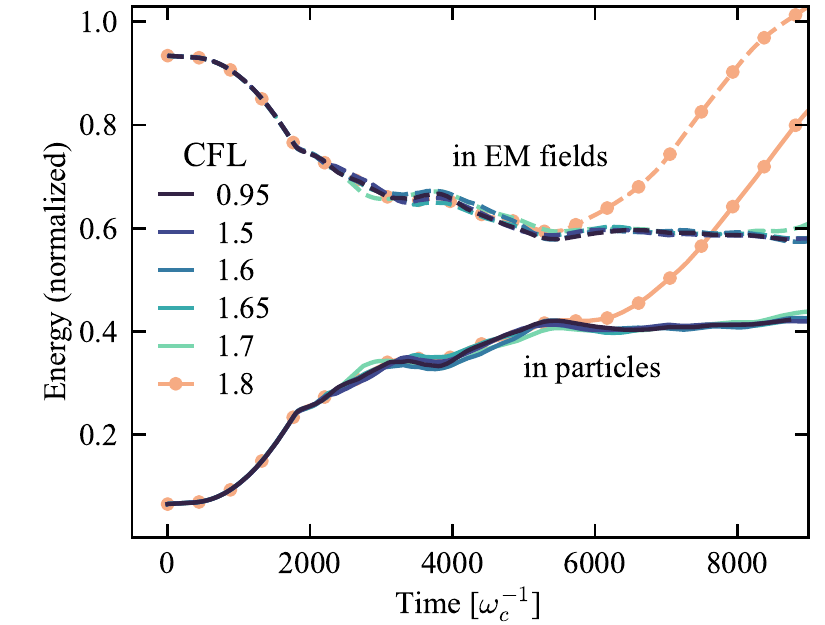}
\caption{\label{fig:energy_total_cfl} Energy balance between electromagnetic field (dashed lines) and particle kinetic energy (solid lines) in PSATD simulations of magnetic reconnection with a CFL factor greater than 1. CFL factors less than $1.7$ produce results that are nearly identical to the benchmark C0.95. For larger values (i.e. C1.8 in orange and C2.0 and C2.2, not shown), we see a qualitative increase in both particle and field energy at progressively earlier times.}
\end{figure}

\begin{figure}
\includegraphics{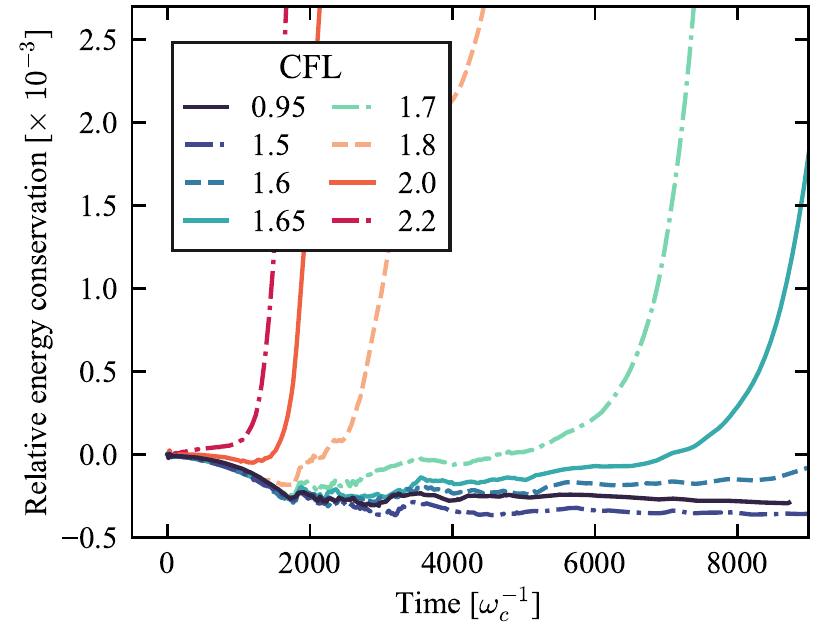}
\caption{\label{fig:energy_conservation_cfl} Relative energy conservation throughout reconnection simulations with CFL factors greater than one. The baseline C0.95 is shown in solid, dark purple. Up to C1.6 (dashed blue), energy non-conservation stays relatively small and comparable to C0.95 (less than one part in 2000). This suggests that a CFL factor $\lesssim 1.6$ is sufficient to capture the necessary physics throughout the time interval of interest ($t \lesssim 9000\,\omegacinv$). For most of the evolution, C1.65 (solid light blue) also keeps a low degree of non-conservation, but larger errors in energy appear at around $t=7000\,\omegacinv$. By the end of the simulation, its non-conservation is still less than 1 percent, but growing rapidly. As we further increase the CFL factor, the rapid increase in energy errors moves to progressively earlier times, with the runaway occurring before $2000\,\omegacinv$ for C2.2 (red dashed-dotted).}
\end{figure}

\begin{figure}
\includegraphics{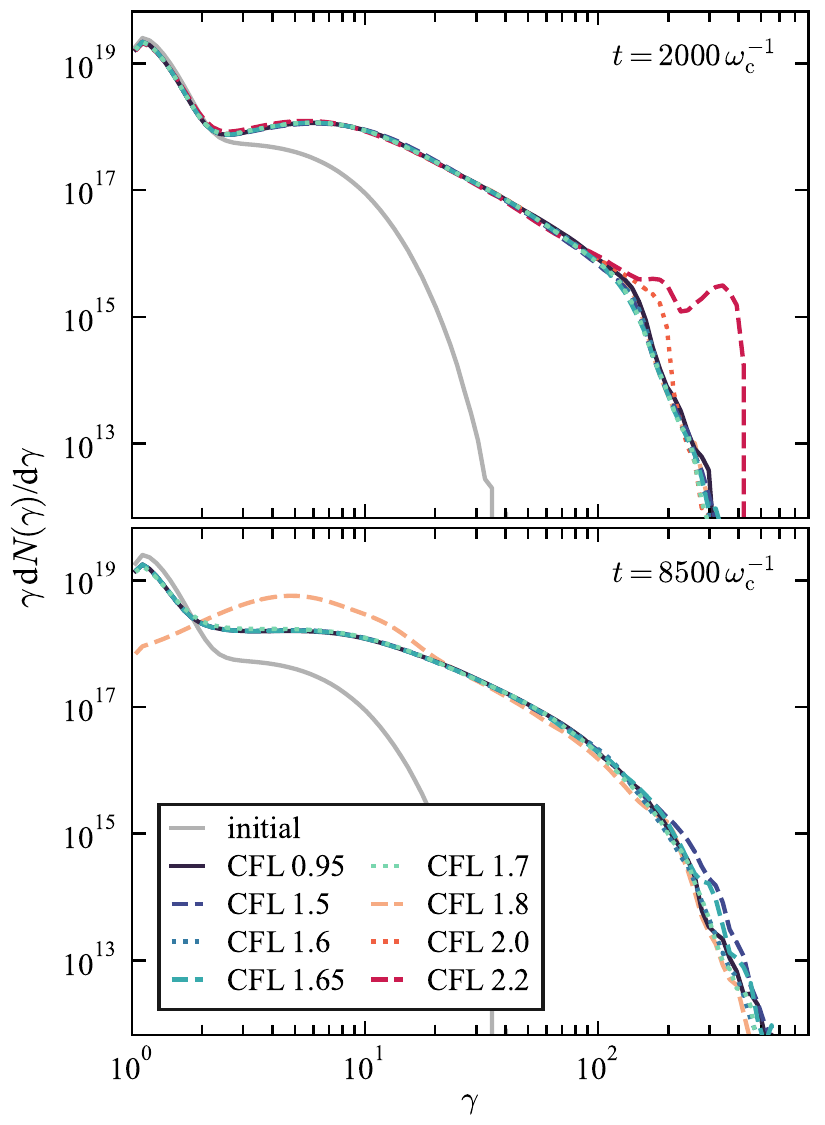}
\caption{\label{fig:power_spectra_cfl} Distribution of particle energy as a function of Lorentz factor ($\gamma$) mid-reconnection (top panel; $t=2000\,\omegacinv$) and near the end of simulations (bottom panel; $t=8500\,\omegacinv$). The baseline C0.95 appears in solid dark purple. Larger values of the CFL factor are shown in dashed and dotted lines. For the values of the CFL factor $\leq 1.7$, the distributions agree closely throughout the simulations, especially at lower energies ($\lesssim 100$). At high energies, there are small differences, which cannot be distinguished from statistical noise. As with energy conservation (Figure~\ref{fig:energy_conservation_cfl}), the simulations with the highest CFLs diverge the earliest; for C2.0 and C2.2 (orange dotted and red dashed), we see disagreement with the baseline simulations at $t=2000\,\omegacinv$, while C1.8 (dashed orange) is still in agreement. By the end of the simulation, C1.8 shows strong disagreement with the baseline, peaking at $\gamma\sim 4$, rather than close to 1. }
\end{figure}

The statistical properties of the reconnecting plasma tend to remain the same early in the simulations and for lower values of the CFL factor. 
For values of the CFL factor $\lesssim 1.65$, the energy conversion from magnetic fields to particles proceeds almost identically (Figure \ref{fig:energy_total_cfl}). We see the same behavior as with the baseline runs (Figure~\ref{fig:energy_total_solver}), where during the initial ${\sim}{1500}\,\omegacinv$, about 20\% of the field energy is transferred to the particles. Following that phase, the energy conversion proceeds more slowly, saturating at a final distribution where about 40\% of the energy is in particles and 60\% remains in the fields.

C1.7 matches the simulations with a lower CFL factor through the main period of interest, until reconnection saturates at around $7000\,\omegacinv$. In the last $1000\,\omegacinv$, it shows a slight increase in both electromagnetic field and particle energy, indicating that energy is not conserved. This is reflected in the relative energy conservation (see Figure~\ref{fig:energy_conservation_cfl}). Non-conservation starts off small but increases sharply at around  $t = 7000\,\omegacinv$, reaching 1\% by the end of the simulation at $ t= 9000\,\omegacinv$.

The runs that all appear the same in the energy balance plot (CFL factor $\leq 1.65$, Figure~\ref{fig:energy_total_cfl}) show greater degrees of energy conservation. For runs with a CFL factor $\lesssim 1.6$, non-conservation is comparable to the baseline, C0.95. C1.65 also shows low levels of non-conservation until the very end of the simulation, when it starts to increase. Left to run further, we expect that non-conservation would continue to increase like in the simulations with higher CFL factors. For the duration of the simulation, C1.65 is within a reasonable tolerance, only violating energy conservation by less than 1 part in 500.

For simulations with CFL factors $\geq 1.8$, significant energy non-conservation develops during the last part of reconnection, much earlier than in C1.7. Errors in the energy have built up significantly by $4000\,\omegacinv$ for C1.8, and by $2000\,\omegacinv$ for both C2.0 and C2.2 (Figure~\ref{fig:energy_conservation_cfl}). These are first apparent in the energy conservation plot, but continue to grow until they show in the plot of energy conversion (orange line in Figure~\ref{fig:energy_total_cfl}). By $t=6000\,\omegacinv$ the field and particle energies are visibly different from the baseline. C2.0 and C2.2 also show earlier runaway growth in field and particle energy, but are omitted from Figure~\ref{fig:energy_total_cfl} for clarity.

The particle energy spectra further demonstrate agreement between the simulations with a CFL factor $\lesssim 1.7$ throughout the majority of the simulation. Figure \ref{fig:power_spectra_cfl} compares the initial particle energy distributions (grey lines) with those at $t=2000\,\omegacinv$ (mid-reconnection) and $t=8500\,\omegacinv$ (after reconnection, almost at the end of the simulation). At the earlier time, all of the runs we study show close agreement with the baseline C0.95 out to $\gamma \sim 100$. Those runs with a CFL factor less than 2 also agree closely with one another at the highest energies; there are few, if any, particles with Lorentz factors in excess of 300. C2.0 shows a slight overabundance of particles with $\gamma\sim 100$, which grows as the simulation progresses. C2.2 shows an even larger overabundance of high-energy particles, reaching a maximum Lorentz factor of $400$, 30\% higher than the maximum achieved by C1.8 at this time. This also coincides with the beginning of the energy non-conservation seen in Figure~\ref{fig:energy_conservation_cfl}. The overabundance of high-energy particles may directly cause the initial non-conservation, but shortly thereafter we also see artificial heating in the cold upstream. In C0.95 and the runs with lower ($\leq1.7$) CFL factors, the cold upstream plasma appears as a peak at $\gamma\sim1$, and remains largely unchanged even at the end of the simulation. However, in C1.8 and above, this peak broadens and moves out to $\gamma\sim4$. This is apparent near the end of C1.8 (bottom panel, Figure~\ref{fig:power_spectra_cfl}). This also occurs in C2.0 and C2.2 beginning shortly after the time of the top panel; we omit their distributions in the later panel for the sake of clarity. 

At the end of the simulations, the runs with CFL factors $\lesssim 1.7$ have particle energy distributions that agree almost exactly for $\gamma<100$. For the highest-energy particles, there are minor differences, comparable to the spread seen in the different electromagnetic solvers (see Figure \ref{fig:power_spectra_solver}). These highest-energy particles are also the rarest -- there are ${\sim} 5000$ times fewer particles at $\gamma=2$ as there are at $\gamma=100$, so the higher energies are more subject to statistical noise.

The non-conservation seen for CFL factors $\geq 1.65$ in Figure~\ref{fig:energy_conservation_cfl} appears to be at least partially driven by numerical heating in the upstream region. The exact cause is uncertain, though we expect it is a direct consequence of the large CFL factor, rather than the under-resolution of a physical timescale. Coincidentally, C1.65 has a time step just over $1\,\omegacinv$, meaning that the cyclotron motion of upstream particles cannot be captured. However, this heating does not occur if we obtain that same time step by, for example, doubling the cell size in each direction and reducing the CFL factor below 1. The effect is driven by the upstream plasma. Simulations of a domain with no current sheet, filled only with upstream ($\theta=0.15$, $\sigma=30$) plasma,   show the same numerical heating.

\begin{figure}
\includegraphics{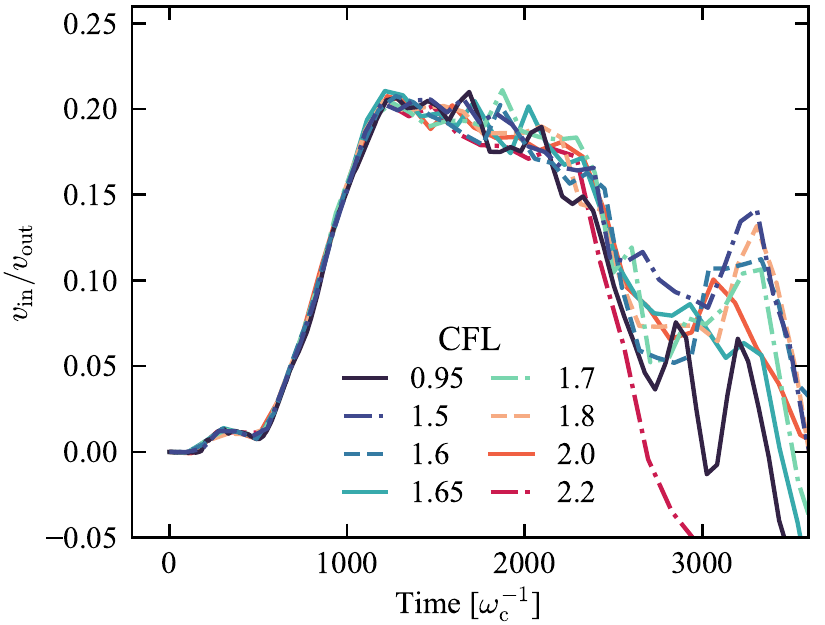}
\caption{\label{fig:recon_rate_cfl} Dimensionless reconnection rate estimated as the ratio of inflow to outflow velocities in the current sheet (see Section~\ref{sec:rate}). The rate appears to be mostly insensitive to the CFL factor. This is largely because our estimate of the reconnection rate is only valid for the first $2000\,\omegacinv$ of the simulations, when energy non-conservation is minimal (Figure~\ref{fig:energy_conservation_cfl}). C2.2 is the only run that shows significant deviation in the first $2500\,\omegacinv$ of evolution, which coincides with a runaway in its energy non-conservation.
}
\end{figure}

The dimensionless reconnection rate is remarkably similar for all runs except for C2.2 (Figure~\ref{fig:recon_rate_cfl}). As discussed in Section~\ref{sec:rate}, we are primarily interested in the value of the estimated reconnection rate for $t<2000\,\omegacinv$. Within that time interval, the simulations all show a nearly-identical rise to $0.2$, followed by a slight decline to 0.15 over the following $1000\,\omegacinv$. C2.2 largely follows this pattern, but declines much more quickly than the other runs at $t=2000\,\omegacinv$. This highlights the robustness of the reconnection rate, and indicates that it likely should not be used to diagnose whether the simulations are producing correct results. At $t=2000\,\omegacinv$, C2.0 and C2.2 both show qualitative disagreement with the baseline models in the high-energy particle spectra (Figure~\ref{fig:power_spectra_cfl}). However, they agree with baseline models in the reconnection rate measurement.

Of the diagnostics discussed here, energy conservation is the most sensitive to numerical problems that arise due to a high CFL factor. Because our simulations are closed systems, the total energy should remain the same throughout the evolution, providing a straightforward ground truth comparison. When increasing the CFL factor, we find that the particle acceleration and reconnection rate obtained from those simulations do not deviate from the baseline until after energy non-conservation begins to rise rapidly. Even C2.2 agrees with C0.95 for the first $1500\,\omegacinv$ of the simulation, before its non-conservation starts to increase. While particle energy spectra may match the baseline even after energy non-conservation begins to run away (e.g.~C1.7 at $t=8500\,\omegacinv$, bottom panel of Figure~\ref{fig:power_spectra_cfl}), we would not be able to verify that particle acceleration was still correct in the absence of a known baseline. When increasing the CFL factor, we suggest using a runaway in energy non-conservation as a heuristic for closed systems to determine when a simulation result is unreliable.

\begin{deluxetable}{cccc}
\tablecaption{Performance comparison of otherwise identical simulations with CFL factors between 0.95 and 2.2. different CFL factors for the PSATD solver and Vay deposition scheme. These numbers are from simulations run to a final time of $t=1470\,\omegac^{-1}$ with dynamic load balancing and no I/O on 21 OLCF Summit nodes (126 GPUs).}
\label{tab:performance_cfl}
\tablehead{CFL & Time step & Walltime & Walltime to \\
Solver & $[\omegacinv]$ & per step [s] & $1470\,\omegac^{-1}$ [s] }
\startdata
0.95 & 0.582 & 0.115 & 290.0\\ 
1.5 & 0.919 & 0.126 & 200.9\\ 
1.6 & 0.978 &  0.127 & 190.9\\ 
1.65 & 1.010 & 0.128 & 186.4\\
1.7 & 1.041 & 0.128 & 180.9\\
1.8 & 1.102 & 0.129 & 172.5\\
2.0 & 1.225 & 0.131 & 157.2\\
2.2 & 1.347 & 0.131 & 143.0\\ 
\enddata
\end{deluxetable}

In Table~\ref{tab:performance_cfl} we summarize the computational performance of simulations with CFL factors greater than 1. Again, we compare against the baseline C0.95 with the PSATD Maxwell solver. We find that between C0.95 and C2.2, the wall time to simulate a fixed physical time interval decreases by half. Over this range of CFL factors, the physical time step increases by a factor of $2.3$, and the walltime per time step increases by about $15\%$. Most of this walltime difference is because current deposition is slower (on a per-step basis) at higher CFL factors. When particles move further per step, they are more likely to move between cells, and this likely reduces the cache performance of the deposition routines.

The speedup obtained for a particular application is limited by how high one can increase the CFL factor while maintaining a reliable result throughout the time interval of interest. We expect that the interval over which energy is conserved for a given CFL factor will vary based on the problem setup. For the configuration described here, a CFL factor of 2.2 may be adequate if we were only interested in the onset of reconnection. In that case, the time to solution will be 50\% of what it is in C0.95.
In this study, we were interested in following reconnection from its onset until it saturates. For that, we needed a CFL factor of at most 1.7, and would have terminated the simulation at around $6500\,\omegacinv$. At that CFL factor, we reach the solution 1.6 times faster than in the baseline C0.95. 

\section{Conclusions}
\label{sec:conclusion}

We have performed first-of-their-kind particle-in-cell simulations of relativistic reconnection with a 16th-order pseudo-spectral Maxwell solver (PSATD) and a time step that exceeds the conventional CFL limit. We find that PSATD and the non-standard finite difference scheme CKC qualitatively and quantitatively produce the same results as the standard second-order finite difference scheme Yee. We have verified that all three schemes produce the same qualitative plasmoid evolution, particle-field energy balance, particle acceleration, and reconnection rate (Figures~\ref{fig:Jy_Bz_1470}--\ref{fig:recon_rate_solver}).
The particle energy distribution has a power law tail with a spectral slope $\alpha$ of $\mathrm{d}N(\gamma)/\mathrm{d}\gamma \propto \gamma^{-\alpha}$. We measure $\alpha$ between $1$ and $3$, as expected \citep{guo:15,werner:16,werner:18}. The reconnection rate is between 0.15 and 0.2 for most of the linear phase of the simulation, consistent with expectations for relativistic reconnection.

We also compare the performance of the solvers and measure efficiency in terms of time to solution. For the same CFL factor, CKC and PSATD allow for a time step that is longer than Yee by a factor of $\sqrt{2}$. The walltime taken per step, though, depends more on the current deposition scheme than on the Maxwell solver. The walltime per step is the same in CKC as in Yee (which both use Esirkepov deposition), giving a ${\sim}40\%$ speedup in time to solution. High-order PSATD is only charge-conserving when used with Vay deposition, which in its current WarpX implementation, is not as optimized as Esirkepov and is therefore more computationally expensive. Thus we decouple the comparison of the solvers and current deposition methods by performing an additional PSATD+Esirkepov simulation. In doing so, we verify that PSATD itself is not the primary cause of the more expensive time step. Despite not being charge-conserving, PSATD+Esirkepov gives the correct answer for this problem, and has a walltime per step only 10\% higher than CKC or Yee. In conjunction with the $\sqrt{2}$-larger time step allowed by PSATD, this produces a net 30\% reduction in time to solution.
A time step in the charge-conserving PSATD+Vay takes 50\% longer than in Yee, so it has a slightly longer time to solution than the most commonly-used charge-conserving Yee algorithm. 

Unlike either of the finite-difference schemes, PSATD is numerically stable at any time step, even one greater than the light travel time across a cell. We explore the accuracy and performance of CFL factors $>1$ in the relativistic reconnection problem. We find that the timescale of interest sets the maximum allowable time step parametrized by the CFL factor. Factors $\leq 1.65$ conserve energy comparably well to the baseline PSATD case until $t=8000\,\omegacinv$, well past the end of reconnection. In that interval, they show good agreement in both particle acceleration (Figure~\ref{fig:power_spectra_solver}) and reconnection rate (Figure~\ref{fig:recon_rate_cfl}). A slightly higher CFL factor of 1.7 still shows good agreement in energy distribution (Figure~\ref{fig:energy_total_cfl}) and particle acceleration, but begins to show signs of runaway errors in energy at $t \approx 6500\,\omegacinv$, right after reconnection saturates. This suggests that for this particular problem and computational setup, the CFL factor could reach 1.7 without compromising the accuracy of the simulation results during reconnection. As we progressively increase the time step (up to a CFL factor of 2.2), runaway energy non-conservation begins earlier and earlier. In the interval where energy is conserved, the other diagnostics such as particle energization and reconnection rate agree with the baseline case, suggesting that energy non-conservation is a good diagnostic of when the simulation results are reliable for closed systems with periodic boundary conditions.

The walltime per time step only increases modestly with CFL factor, about 15\% from the baseline C0.95 to C2.2. The increases in the time step outweigh the increases in walltime per step, reducing the walltime per physical time by a factor of about two between C0.95 and C2.2, if all other algorithmic choices are unchanged. Comparing against CKC+Esirkepov, the most efficient of the CFL=0.95 simulations, we obtain a 25\% reduction in time to solution by adopting a CFL factor of 2.2, and a $<10\%$ reduction by using the CFL factor of 1.7 that we have determined to be suitable for our scientific purposes. If a CFL factor below 1.7 were necessary, CKC+Esirkepov would be the most efficient option due to the faster current deposition. Future work will include further optimization of the Vay deposition routine, which would shift the trade-offs to favor the PSATD+Vay combination in more situations. While PSATD+Esirkepov was accurate in these simulations, we do not recommend its use without an additional current correction or divergence-cleaning operation to guarantee charge conservation.

One of PSATD's strengths is that it reduces numerical dispersion that may appear when using Yee or CKC solvers. While an exploration into the mitigating effect of the PSATD solver on numerical dispersion in relativistic plasmas is outside the scope of this paper, it is indeed a known effect that can contaminate simulations of astrophysical jets, shocks, and magnetic reconnection \citep{godfrey:74, melzani:13, godfrey:14, ikeya:15, li:17, nishikawa:21, tomita:22}. PSATD can therefore provide new opportunities to study highly-relativistic plasmas, a numerically challenging regime which characterizes a number of astrophysical systems.

The numerical and algorithmic innovations leveraged in this study can be used to enable larger and more efficient simulations of astrophysical and laboratory plasmas \citep{ji:22}.
Our simulations are also some of the first GPU-accelerated astrophysical PIC simulations. As a first step, we have only focused on two-dimensional systems without additional kinetic and radiative physics. A third spatial dimension is dynamically important in reconnection because it makes the current sheet susceptible to an additional instability, called ``drift-kink'', which can suppress particle acceleration \citep[e.g.][]{cerutti:14_recon_rr,sironi:14,guo:15,werner:17}. CKC and PSATD may be especially efficient in 3D because their time steps are larger than Yee's by a factor of $\sqrt{3}$, rather than $\sqrt{2}$ in 2D. In many astrophysical reconnection environments, synchrotron emission and pair production play an important role. With WarpX, we will be able to take advantage of GPU-accelerated exascale computing resources to perform 3D simulations that include this radiative physics. 
\\
\\
%\begin{acknowledgments}
We thank Luca Comisso for the insightful discussions on magnetic reconnection and reconnection rate.
This work was partially supported by the U.S.~Department of Energy, Office of Science, Office of Advanced Scientific Computing Research, Exascale Computing Project (17-SC-20-SC) under contract DE-AC02-05CH11231, and Simulation and Analysis of Reacting Flow, FWP \#FP00011940, funded by the Applied Mathematics Program in ASCR. 
This research used the open-source particle-in-cell code WarpX \url{https://github.com/ECP-WarpX/WarpX}, primarily funded by the US DOE Exascale Computing Project.
Primary WarpX contributors are with LBNL, LLNL, CEA-LIDYL, SLAC, DESY, CERN, and Modern Electron.
We acknowledge all WarpX contributors. 
This work used resources of the Oak Ridge Leadership Computing Facility at the Oak Ridge National Laboratory, which is supported by the Office of Science of the U.S. Department of Energy under Contract No. DE-AC05-00OR22725. 
This research also used resources of the National Energy Research Scientific Computing Center (NERSC), a U.S. Department of Energy Office of Science User Facility located at Lawrence Berkeley National Laboratory, operated under Contract No. DE-AC02-05CH11231 using NERSC awards DDR-ERCAP m3881 for 2022 and ASCR-ERCAP mp111 for 2022 and 2023.
%\end{acknowledgments}

\defcitealias{warpx_zenodo}{WarpX Development Team 2023}
\software{
WarpX \citepalias{fedeli:22,warpx_zenodo},
AMReX \citep{zhang:19,amrex_zenodo},
matplotlib \citep{matplotlib},  
numpy \citep{numpy},
scipy \citep{scipy},
yt \citep{yt}
}

\appendix
\section{Magnetic Field Configuration}
\label{app:magnetic_field}

\subsection{Equilibrium}

\begin{deluxetable}{ccc}
\tablecaption{Physical parameters and symbols common to all of our simulations. \label{tab:params_numerical}}
\tablehead{Parameter & Symbol & Value}
\startdata
Background Larmor radius & $\rhoc$ & $4.1\times10^{-3}\,\mathrm{m}$\\
Background Larmor frequency & $\omegac$ & $7.3\times10^{-10}\,\mathrm{s}^{-1}$\\
Skin depth & $\lambdae$ & $0.01 \,\mathrm{m}$\\
Current sheet half-width & $\delta$ & $0.05 \,\mathrm{m}$\\
Background magnetization & $\sigma$ & 30 \\
Background magnetic field & $B_0$ & $0.42 \,\mathrm{T}$\\
Current sheet number density & $\nd$ & $2.8 \times 10^{17} \,\mathrm{m}^{-3}$\\
Background number density & $\nb$ & $5.6 \times 10^{16} \,\mathrm{m}^{-3}$\\
Current sheet velocity & $\beta_0$ & $0.22\, c$  \\
Background temperature & $\theta_\mathrm{b}$ & $0.15$ \\
Current sheet temperature & $\theta_\mathrm{d}$ & $1.57$ \\
Domain half-width (x) & $L_x$ & $8.96\,\mathrm{m}$\\
Domain half-width (z) & $L_z$ & $4.32\,\mathrm{m}$\\
\enddata
\end{deluxetable}

In all of our simulations, we fix $\rhoc = 4.1\times 10^{-3}\,\mathrm{m}$, which gives $B=0.42\,\mathrm{T}$ and fixes the other dimensional values in Table~\ref{tab:params_numerical}. These are used in the WarpX simulations, which require SI units. This choice of length scale is largely arbitrary. In this non-radiative regime, the magnetization determines most of the physics; the rest of the results will scale accordingly.

From the relationship between $\rhoc$ and $\lambdae$ and the definition of skin depth, we can calculate number density in the current sheet $\nd$:
\begin{equation}
    \nd = \frac{m \epsilon_0 c^2}{e^2\lambdae^2} = 2.8 \times 10^{17} \,\mathrm{m}^{-3}
\end{equation}
and in turn background number density $\nb = 5.6 \times 10^{16}\,\mathrm{m}^{-3}$. At this density, a magnetization of 30 requires the background magnetic field $B_0 = 0.42\,\mathrm{T}$.

After integrating Ampere's Law (equation~\ref{eq:amplaw}),
\begin{align}
    B_z(x) =& -4\mu_0 e \nb c \beta_0 \delta \left[\arctan\left(\tanh\left(\frac{x+\xc}{2\delta}\right)\right)-\arctan\left(\tanh\left(\frac{x-\xc}{2\delta}\right)\right)\right]\nonumber\\&+\frac{1}{2}\left(\frac{\nd}{\nb}-1\right)\left(	\tanh\left(\frac{x+\xc}{\delta}\right)-	\tanh\left(\frac{x-\xc}{\delta}\right)\right)+C.
\end{align}
To solve for the constant of integration $C$, we apply the condition from equation \ref{eq:bcs}:
\begin{align}
    B_z(\infty) = -B_z(0) = B_0.
\end{align}
First taking the limit as $x\rightarrow\infty$,
\begin{align}
\lim_{x\rightarrow\infty}B_z(x) &= - 4 \mu_0 e \nb \beta_0 c \delta \left( \arctan\left(1\right)- \arctan\left(1\right) + \frac{1}{2}\left(\frac{\nd}{\nb}-1\right)\left( 1- 1 \right)\right)\nonumber + C\\
 &= C,
\end{align}
since $\lim_{x\rightarrow\infty} \tanh(x) = 1$.
To double precision, $\tanh(x) = 1$ if $x\geq20$, so the following is true if $\xc/\delta \geq 40$, which is the case in our simulations:
\begin{align}
B_z(0) &= - 4 \mu_0 e \nb \beta_0 c \delta \left(\arctan\left(1\right)-\arctan\left(-1\right) + \frac{1}{2}\left(\frac{\nd}{\nb}-1\right)\left( 1 - (-1)\right)\right)\nonumber + C\\
 &= - 4 \mu_0 e \nb \beta_0 c \delta \left(\frac{\pi}{2} + \frac{\nd}{\nb}-1\right) + C.\\
\end{align}

Combining the conditions on $B_z$ as $x\rightarrow\infty$ and at $x=0$ (equation \ref{eq:bcs}):
\begin{align}
   B_0 = \lim_{x\rightarrow\infty} B_z(x) = C &= -B_z(0)\\
   C &= 4\mu_0 e \nb \beta_0 c \delta \left(\frac{\pi}{2}+\frac{\nd}{\nb}-1\right) - C\\
   C &= 2\mu_0 e \nb \beta_0 c \delta \left(\frac{\pi}{2}+\frac{\nd}{\nb}-1\right).
\end{align}

This yields the expression for magnetic field in equations~\ref{eq:bz} and \ref{eq:B0}:
\begin{align}
    B_z(x) = - \frac{ 2 B_0 }{\left(\frac{\pi}{2} + \frac{\nd}{\nb}-1\right)} &\left( \arctan\left(	\tanh\left(\frac{x+\xc}{2\delta}\right)\right)- \arctan\left(\tanh\left(\frac{x-\xc}{2\delta}\right)\right) - \frac{\pi}{4} \right. \nonumber\\
    &+ \left.\frac{1}{2}\left(\frac{\nd}{\nb}-1\right)\left(\tanh\left(\frac{x+\xc}{\delta}\right)- 	\tanh\left(\frac{x-\xc}{\delta}\right) - 1\right)\right),
\end{align}
with 
\begin{align}
   B_0 &= 2\mu_0 e \nb \beta_0 c \delta \left(\frac{\pi}{2}+\frac{\nd}{\nb}-1\right).
\end{align}

The full expressions for the equilibrium fields and plasma properties are:
\begin{align}
    \frac{B(x)}{B_0} = \frac{2}{\frac{\pi}{2} + \frac{\nd}{\nb}-1} &\left[\arctan\left(\tanh\left(\frac{x-\xc}{2\delta}\right)\right)
    - \arctan\left(\tanh\left(\frac{x+\xc}{2\delta}\right)\right) + \frac{\pi}{4}\right.\nonumber\\
    &+ \left.\frac{1}{2}\left(\frac{\nd}{\nb}-1\right) \left(\tanh\left(\frac{x-\xc}{\delta}\right) - \tanh\left(\frac{x+\xc}{\delta}\right) + 1\right)
    \right] \label{eq:Bx_full}
\end{align}
\begin{equation}
    \frac{n(x)}{\nb} =  1 + \left(\frac{\nd}{\nb}-1\right)\left(\sech\left(\frac{x+\xc}{\delta}\right) + 
    \sech\left(\frac{x-\xc}{\delta}\right)\right) \label{eq:nx_full}
\end{equation}
\begin{align}
    \beta(x) &=\frac{\lambdae \sqrt{\sigma}}{2\delta\left(\frac{\pi}{2} + \frac{\nd}{\nb}-1\right)} \sqrt{\frac{\nd}{\nb}}\left(\sech\left(\frac{x+\xc}{\delta}\right) - \sech\left(\frac{x-\xc}{\delta}\right)\right) \label{eq:betax_full}
\end{align}
\begin{equation}
    \theta(x) = \frac{\sigma}{4} \frac{(4+\eta)/\eta- [B_z(x)/B_0]^2}{[n(x)/\nb]\sqrt{1-\beta(x)^2}}.
\end{equation}

\subsection{Magnetic Field Perturbation}

The magnetic field perturbation is based on the the vector potential (equation~\ref{eq:Ay}), which is an integral of $B_z(x)$ (equation \ref{eq:bz}).
We split $B(x)$ into three types of terms, based on functional form: constants, $\arctan(\tanh)$ terms, and $\tanh$ terms. Constants are straightforward to integrate. The integrals of the other two types of terms are:
\begin{align}
\int \tanh&\left({\frac{x\pm \xc}{\delta}}\right)\,\mathrm{d}x \equiv \zeta_\pm(x) = \delta\log\cosh\left(\frac{x\pm \xc}{\delta}\right)\\
\int \arctan\tanh&\left(\frac{x\pm \xc}{2\delta}\right) \,\mathrm{d}x \equiv \xi_\pm(x)\nonumber\\
&= 2 \delta \left(\frac{x\pm \xc}{2\delta}\right)\left[\arctan\left(\exp\left(-\frac{x\pm \xc}{\delta}\right)\right) + \arctan\left(\tanh\left(\frac{x\pm \xc}{2 \delta}\right)\right)\right]\nonumber\\
&+\frac{\delta i}{2} \left[\dilog\left(-i \exp\left(-\frac{x\pm \xc}{\delta}\right) \right) - \dilog\left(i \exp\left({-\frac{x\pm \xc}{\delta}}\right)\right)\right]. 
\end{align}

The value of $\int B_z(x) \,\mathrm{d}x$ is:
\begin{align}
    \int B_z(x) \,\mathrm{d}x = \frac{2 B_0}{\frac{\pi}{2} + \frac{\nd}{\nb} - 1} & \left[\xi_-(x) - \xi_+(x) + \frac{\pi}{4} x + \frac{1}{2}\left(\frac{\nd}{\nb}-1\right) \left(\zeta_-(x) - \zeta_+(x) + x\right) \right]\\
    = \frac{2 B_0}{\frac{\pi}{2} + \frac{\nd}{\nb} - 1} & \left[\frac{\pi}{4} x + (x-\xc)\left(\arctan\exp\left(-\frac{x-\xc}{\delta}\right) + \arctan\tanh\left(\frac{x-\xc}{2\delta}\right)\right)\right.\nonumber\\
    &-\left.(x+\xc)\left(\arctan\exp\left(-\frac{x+\xc}{\delta}\right) + \arctan\tanh\left(\frac{x+\xc}{2\delta}\right)\right)\right.\nonumber\\
    &+ \left.\frac{\delta i}{2}\left(\dilog\left(-i \exp\left(-\frac{x-\xc}{\delta}\right)\right)-\dilog\left(i\exp\left(-\frac{x-\xc}{\delta}\right)\right)\right)\right.\nonumber\\
    &-\left.\frac{\delta i}{2}\left(\dilog\left(-i \exp\left(-\frac{x+\xc}{\delta}\right)\right)-\dilog\left(i\exp\left(-\frac{x+\xc}{\delta}\right)\right)\right)\right.\nonumber\\
    &+\left.\frac{1}{2}\left(\frac{\nd}{\nb}-1\right)\left(\delta \log\cosh\left(\frac{x-\xc}{\delta}\right) - \delta\log\cosh\left(\frac{x+\xc}{\delta}\right)+x\right)\right]\\
    = \frac{2 B_0}{\frac{\pi}{2} + \frac{\nd}{\nb} - 1} & \left[\frac{\pi}{4} (x-2\xc) + \frac{1}{2}\left(\frac{\nd}{\nb}-1\right)\left(\delta \log\cosh\left(\frac{x-\xc}{\delta}\right) - \delta\log\cosh\left(\frac{x+\xc}{\delta}\right)+x\right)\right.\nonumber\\
    &+ \frac{\delta i}{2}\left(\dilog\left(-i \exp\left(-\frac{x-\xc}{\delta}\right)\right)-\dilog\left(i\exp\left(-\frac{x-\xc}{\delta}\right)\right)\right)\nonumber\\
    &-\left.\frac{\delta i}{2}\left(\dilog\left(-i \exp\left(-\frac{x+\xc}{\delta}\right)\right)-\dilog\left(i\exp\left(-\frac{x+\xc}{\delta}\right)\right)\right)\right]. \label{eq:A_wo_arctan}
\end{align}
Equation \ref{eq:A_wo_arctan} results from applying the simplification
\begin{equation}
    \arctan\left(e^{-2y}\right)+ \arctan\left(\tanh(y)\right) = \frac{\pi}{4}.
    \label{eq:arctan_simplification}
\end{equation}
For the purposes of symmetry and preserving the periodic boundary conditions, we have set the constant of integration equal to zero.

The perturbation to the magnetic field is:
\begin{align}
\vect{B'} &= \vect{\nabla}\times A_y - \vec{B} =  -\frac{\partial A_y}{\partial z} \mathbf{\hat{i}} + \frac{\partial A_y}{\partial x} \mathbf{\hat{k}} - \vec{B}\\
&=-0.01 \times \mathbf{\hat{i}}\left[\frac{51 \pi}{L_z} \sin\left(\frac{\pi}{L_z} z\right) \cos^{50}\left(\frac{\pi}{L_z} z\right)\cos^2 \left(\frac{\pi (x-\xc)}{L_x}\right) \int B_z(x) \,\mathrm{d}x\right] +\nonumber\\
&+ 0.01\times \mathbf{\hat{k}}\left[ \frac{\pi}{L_x}\cos^{51}\left(\frac{\pi}{L_z}z\right)\sin\left(\frac{2\pi (x-\xc)}{L_x}\right) \int B_z(x)\,\mathrm{d}x - \cos^{51}\left(\frac{\pi}{L_z}z\right) \cos^2\left(\frac{\pi(x-\xc)}{L_x}\right) B_z(x)\right]. \label{eq:full_perturbation}
\end{align}

\bibliography{auto_generated_bibliography,manual_bib}{}
\bibliographystyle{aasjournal}

\end{document}